# Non-volatile optical phase shift in ferroelectric hafnium zirconium oxide

Kazuma Taki[1], Naoki Sekine[1], Kouhei Watanabe[1], Yuto Miyatake[1], Tomohiro Akazawa[1], Hiroya Sakumoto[1], Kasidit Toprasertpong[1], Shinichi Takagi[1], and Mitsuru Takenaka[1*]

[1] Department of Electrical Engineering and Information Systems,

The University of Tokyo, 7-3-1 Hongo, Bunkyo-ku, Tokyo 113-8656,

Japan, Phone: +81-3-5841-6733, Fax: +81-3-5841-8564,

*E-mail: takenaka@mosfet.t.u-tokyo.ac.jp

**Abstract**

A non-volatile optical phase shifter is a critical component for enabling large-scale, energy-efficient programmable photonic integrated circuits (PICs) on a silicon (Si) photonics platform, facilitating communication, computing, and sensing[1]. While ferroelectric materials like $BaTiO_3$ offer non-volatile optical phase shift capabilities[2], their compatibility with complementary metal-oxide-semiconductor (CMOS) fabs is limited. Hence, the search for a novel CMOS-compatible ferroelectric material for non-volatile optical phase shifting in Si photonics is of utmost importance. Hafnium zirconium oxide (HZO) is an emerging ferroelectric material discovered in 2011[3-5], which exhibits CMOS compatibility due to the utilization of high-k dielectric $HfO_2$ in CMOS transistors. Although extensively studied for ferroelectric transistors and memories[6,7], its application in photonics remains relatively unexplored[8-10]. Here, we show the optical phase shift induced by ferroelectric HZO deposited on a SiN optical waveguide. We observed a negative change in refractive index at a 1.55 μm wavelength in the pristine device regardless of the direction of an applied electric filed. We achieved approximately π phase shift in a 4.5-mm-long device with negligible optical loss. The non-volatile multi-level optical phase shift was confirmed with a persistence of > $10^4$ s. This phase shift can be attributed to the spontaneous polarization within the HZO film along the external electric field. Our results pave the way for large-scale integration of optical phase shifters on a 300-mm-diameter Si wafer, utilizing a CMOS-compatible process. We anticipate that our results will stimulate further research on optical nonlinear effects, such as the Pockels effect, in ferroelectric HZO. This advancement will enable the development of various devices, including high-speed optical modulators. Consequently, HZO-based programmable PICs are poised to become indispensable in diverse applications, ranging from optical fiber communication and artificial intelligence to quantum computing and sensing.

## Introduction

Silicon (Si) photonics have experienced remarkable advancements over the past two decades, playing a pivotal role in facilitating the exponential growth of optical fiber communication traffic[11,12]. This progress has led to an exponential increase in the integration capacity of optical elements on a Si photonic integrated circuit (PIC), fueling expectations for the development of programmable PICs, catering not only to optical communications but also to computing and sensing applications[13-17]. An essential component in reconfiguring the functionality of a programmable PIC is the optical phase shifter. This component plays a crucial role in adjusting the phase of the light signal as it propagates through an optical waveguide. In the context of a Si programmable PIC, the thermo-optic (TO) phase shifter is the most commonly used type of optical phase shifter. While the TO phase shifter has a simple structure and can be easily integrated, its drawback lies in its large power consumption, which poses challenges for large-scale integration[18]. However, alternative approaches have been explored to address this issue. Low-power and low-loss optical phase shifters have been reported using III-V/Si hybrid integration[19], micro-electro-mechanical systems (MEMS)[20], ferroelectrics[21-23], phase change materials[24,25], 2D materials[26,27], and piezoelectric actuation[28]. Among these alternatives, the optical phase shifter based on ferroelectric $BaTiO_3$ shows promise, particularly due to its low-loss, non-volatile operation driven by an external electric field[2,21]. The non-volatile nature of the $BaTiO_3$-based phase shifter allows for zero static power consumption and simplification of electric wiring through a crossbar configuration[29]. Moreover, the non-volatility of a phase shifter is essential for an energy-efficient deep learning accelerator based on a programmable PIC utilizing an in-memory computing architecture, which helps overcome the von Neumann bottleneck[30]. However, a challenge arises with $BaTiO_3$, as it is incompatible with the complementary metal-oxide-semiconductor (CMOS) process, making it difficult to fabricate $BaTiO_3$-based optical phase shifters in CMOS fabs. Hence, there is a strong desire for a CMOS-compatible ferroelectric material.

In this work, we investigate the non-volatile optical phase shift in orthorhombic $Hf_{0.5}Zr_{0.5}O_2$ (HZO) which was discovered to exhibit ferroelectricity in 2011[3-5]. Since the $HfO_2$-based ferroelectrics are CMOS compatible, a wide variety of electron devices such as ferroelectric transistors and ferroelectric memories have intensely been investigated[6,7]. However, the optical properties of the $HfO_2$-based ferroelectrics have been not fully explored yet[8-10]. Therefore, it is of utmost importance to evaluate the refractive index change induced by the electric field in ferroelectric HZO, as it has the potential to be applied as a CMOS-compatible Si optical phase shifter.

## Device structure and fabrication

To evaluate the electric field induced refractive index change in HZO, we prepared a SiN optical waveguide surrounded by HZO (10 nm)/$Al_2O_3$ (1 nm) stacks (See Method and Extended Data Fig. 1a and 1b) with a Mach-Zehnder interferometer (MZI), as shown in Figs. 1a and 1b. The SiN waveguide operating at a 1.55 μm wavelength was fabricated on a thermally oxidized Si wafer. It is known that the ferroelectricity in HZO reaches its maximum when crystalizing at the film thickness of 10 nm[31]. To enhance the overlap between the confined light in the waveguide and the HZO layers, we stacked 10-nm-thick HZO films in three cycles with 1-nm-thick $Al_2O_3$ interlayers using atomic layer deposition (ALD) so that the total ferroelectric thickness achieves 30 nm while each 10-nm-thick layer is separately crystalized. The $Al_2O_3$ interlayer induces oxygen vacancies into the HZO layers, thereby promoting the HZO layers to exhibit ferroelectric behavior[32]. A bias voltage was applied to generate a horizontal electric field across the waveguide, as displayed in Fig. 1b. Figure 1c shows a plan view of the fabricated SiN waveguide devices. We prepared an asymmetric Mach-Zehnder interferometer (AMZI) with 4.5-mm-long phase modulators, enabling us to evaluate the phase shift through the resonance wavelength peak shift in its transmission spectrum. To evaluate the change in the optical loss with respect to the application of a bias voltage, we also prepared straight optical phase modulators. We also fabricated the samples with nonferroelectric $HfO_2$ or $SiO_2$ cladding layer as references (See Extended Data Fig. 1c and 1d). The width of the SiN waveguide was designed to be 1.2 μm for a single-mode operation (See Extended Data

Fig. 2). The multimode interference couplers employed in AMZIs and the grating couplers for fiber coupling were designed for the transverse-electrical (TE) mode of the SiN waveguide (See Extended Data Fig. 2e and 2f).

Figure 1d presents a cross-sectional transmission electron microscopy (TEM) image of the SiN waveguide with the conformally deposited HZO/$Al_2O_3$ layers. A magnified TEM images clearly revealed three polycrystalline HZO layers separated by $Al_2O_3$ interlayers (See Extended Data Fig. 3). Automated crystal orientation mapping in transmission electron microscopy (ACOM-TEM)[33] with a precession electron diffraction angle of 0.5° was conducted to analyze the phase and orientation of the polycrystalline HZO layers, as depicted in Fig. 1e. We found that the majority of the HZO layers on the top and sidewalls of the SiN waveguide were in the orthorhombic crystal phase, which exhibits ferroelectric properties.

## Ferroelectric properties

First, we evaluated the properties of the HZO/$Al_2O_3$ stacks prepared on a Si substrate, as shown in Fig. 2. To evaluate the crystallinity of the poly HZO/$Al_2O_3$ stacks, we performed grazing-incidence X-ray diffraction as shown in Fig. 2a. We found that the diffraction peaks correspond to the orthorhombic phases[5]. Polarization−voltage (PV) characteristics of the metal-insulator-semiconductor capacitor composed of the HZO/$Al_2O_3$ stacks with varied voltage ranges were shown in Fig. 2b. As expected to the measured orthorhombic phase in Fig. 1d and Fig. 2a, the clear hysteresis PV was observed, verifying the ferroelectricity of the HZO/$Al_2O_3$ stacks. When a bias voltage was swept between -15 V to +15 V, the remnant polarization of the HZO/$Al_2O_3$ stacks was approximately 20 $\mu C/cm^2$, comparable to that of a single-layer, 10-nm-thick HZO[34,35]. By inserting a 1 nm $Al_2O_3$ layer for every 10 nm of the HZO layer[36,37], we achieved good ferroelectricity even in the HZO layer with a total thickness of 30 nm (See Extended Data Fig. 4).

**Optical properties**

Next, we measure the transmission of the fabricated waveguide devices to evaluate the refractive index change of the HZO layer in response to an applied voltage. The measurement was conducted by injecting a continuous-wave laser light with wavelengths ranging from 1480 nm to 1580 nm through grating couplers (see Method). As shown in Extended Data Fig. 5, the propagation loss of the SiN waveguide was evaluated to be 6.1 dB/cm. Since the thermo-optic (TO) effect could potentially contribute to the change in the effective refractive index of the propagating light, we conducted measurements of the leakage current when applying a bias voltage between the two electrodes of the phase shifter. Our findings revealed a leakage current of less than approximately 300 pA with a 200 V bias voltage (See Extended Data Fig. 6). Considering that the TO effect typically requires more than 10 mW to induce a $\pi$ phase shift in a SiN waveguide[38], we can confidently conclude that the TO effect is negligible in our experiments. We also investigated the impact of free-carrier plasma dispersion and absorption induced by an applied electric field, which is particularly significant in the case of Si. This examination involved measuring the transmission of a 10-mm-long straight waveguide in response to the application of a bias voltage, as shown in Fig. 3a. Notably, we observed no optical intensity modulation even with the application of a 210 V bias voltage, suggesting that the free-carrier effect in the SiN waveguide is negligible.

Then, we proceeded to directly assess the optical phase shift using the AMZI

configuration by applying a bias voltage to one of the MZI arms, as illustrated in Fig. 3b. The transmission spectrum of the AMZI before bias voltage application is represented by the black line in Fig. 3b. Notably, the spectrum exhibits periodic wavelength resonance peaks attributed to the length difference between the two MZI arms. Subsequently, we applied a bias voltage of 210 V to a 4.5-mm-long optical phase shifter within the longer AMZI arm, as indicated by the red line in Fig. 3b. The direction of the applied electric field is illustrated as a green arrow in the inset of Fig. 3. The transmission spectrum in this case reveals a distinctive blue shift in the resonance wavelength peak, clearly indicating a negative change in the effective refractive index of the propagating light. Importantly, the blue-shifted wavelength peak remained unchanged even upon returning the bias voltage to 0 V, indicating the attainment of a non-volatile optical phase shift. The change in the effective refractive index of the pristine HZO device in response to an applied voltage is depicted by the green circular symbols in Fig. 3c. In this measurement, a bias voltage was applied to the shorter AMZI arm, as depicted in the inset of Fig. 3c. Although no notable change in the effective refractive index occurred within the range of applied voltages from 0 V to 120 V, a substantial negative change in the effective refractive index became evident when the applied voltage exceeded 120 V. Upon reverting the applied voltage from 210 V back to 0 V, the change in the effective refractive index remained relatively constant, as depicted in Fig. 3b, with a magnitude of approximately $-1.5\times10^{-4}$ even at 0 V. In this measurement, the direction of the electric field is opposite to that in Fig. 3b, yet the refractive index change remains negative, similar to Fig. 3b. Therefore, the initial substantial change in the refractive index is consistently negative, irrespective of the electric field's direction. Continuing to sweep the voltage down to -210 V did not exhibit the same large change in the effective refractive index that was initially observed. It is worth highlighting that the $HfO_2$ and $SiO_2$ reference devices exhibited a positive change in the effective refractive index, which is approximately 10 times smaller than the HZO device, as shown in the black line of Fig. 3c and Extended Data Fig. 7. The refractive index change increased with the square of the voltage, which can be attributed to the Kerr effect in SiN[39]. Based on this comparison, we can deduce that the substantial negative change in the effective refractive index originates from the ferroelectric HZO layer. The retention characteristics of the non-volatile refractive index changes observed in the HZO sample were then evaluated. Figure 4d shows the results of measuring the change in refractive index over time after a voltage of 120 V or 160 V was applied and then the voltage was returned to 0 V again. It can be seen that there is no significant change in the refractive index throughout measurements up to 10000 s. This result strongly suggests the potential realization of a multi-level non-volatile optical phase shifter using ferroelectric HZO.

## Discussion and outlook

In order to delve into the underlying physical mechanism behind the non-volatile optical phase shift in ferroelectric HZO, we conducted ACOM-TEM on the HZO layer. This analysis allowed us to explore the alteration in the orientation of the orthorhombic phase induced by the application of an electrical field. Figure 4a illustrates a grain map of the orthorhombic HZO layer on the SiN waveguide before the application of any electrical field. In Fig. 4b, an inverse pole figure of the HZO layer along the direction of the electrical field shown in Fig. 1a, defined as the transverse direction, is presented; here, the (001) orientation is defined as the polarization axis of orthorhombic HZO crystal. Figures 4c and 4d show a grain map and an inverse pole figure of the HZO layer, respectively, after the application of 210 V. By comparing Fig. 4b and Fig. 4d, it becomes evident that the polarization axis shifts more toward the transverse direction due to the applied voltage[40,41]. Figure 4e shows the product of the normalized distribution density of the orientation angle θ of the orthorhombic phase and cosθ, which is regarded as the component contributing to the change in refractive index change. Here, θ is defined as an angular between the polarization axis and the transverse direction. It is clearly seen that the product-sum of the normalized distribution density and cosθ increased by approximately 30% after the voltage application. We observed the similar trend in the HZO layer at the sidewall of the SiN waveguide, as shown in Extended Data Fig. 8.

Taking into account the rotation of the polarization axis of HZO along the direction of the applied electric field, as observed in Fig. 4, the negative change in the refractive index, as depicted in Fig. 3c, regardless of the electric field's direction, can be elucidated through a couple of physical models. One such model is the quadratic electro-optic effect induced by spontaneous polarization of HZO[42-44]. As observed in $LiNbO_3$, the refractive index reduction is proportional to the square of the spontaneous polarization through the quadratic electro-optic effect in ferroelectric materials. When a voltage is applied to the pristine device, the spontaneous polarization is expected to evolve along the transverse direction, as depicted in Fig. 4. As a result, we expect to observe a negative refractive index change that is independent of the direction of voltage application. The second model pertains to the birefringence of HZO. According to numerical predictions, the refractive index of orthorhombic $HfO_2$ along the polarization axis is 2.081 at a wavelength of 633 nm, while the refractive indices along the other two axes are anticipated to be 2.059 and 2.122[10]. Even at a wavelength of 1550 nm, the refractive indices are expected to be close to those at 633 nm[45]. In directions perpendicular to the polarization axis, HZO crystals are assumed to be randomly oriented. Consider the scenario where the polarization axis is oriented perpendicular to the transverse direction. Since the measurement in Fig. 3 employs TE-mode light, where the optical electric field is predominantly polarized along the transverse direction, the refractive index for the propagating light becomes the average of the refractive index perpendicular to the polarization axis, i.e., 2.0905. Upon rotating the polarization axis along the transverse direction through voltage application, the refractive index for the propagating light changes to the value along the polarization axis, i.e., 2.081. The same logic applies to $ZrO_2$. Consequently, the refractive index reduction observed regardless of the direction of voltage application is attributed to the birefringence of orthorhombic HZO.

In conclusion, we successfully demonstrated the non-volatile optical phase shift induced by applying a voltage to ferroelectric HZO deposited on a SiN waveguide. This observed phenomenon can be attributed to the rotation of the polarization axis of orthorhombic HZO along the direction of the external electric field. The non-volatile optical phase shift observed in HZO holds significant utility for reconfiguring programmable PICs, as well as for adjusting initial optical phase errors that are inevitably caused by fabrication variations. Presently, due to limitations in the measurement equipment, applying a voltage higher than 210 V is not feasible. Nevertheless, it is anticipated that employing a larger electric field on HZO could yield an even more substantial optical phase shift. In the present paper, the linear electro-optic effect, the Pockels effect has not been observed in HZO. However, there is a possibility of observing the Pockels effect in HZO by applying a larger electric field using a slot waveguide or plasmonic waveguide. We anticipate that that our paper will trigger and promote the study of optical nonlinear effects in HZO, leading to the progress of Si PICs employing ferroelectric HZO and making significant contributions to the evolution of communications, computing, and sensing.

## Methods

The SiN waveguide with the HZO/$Al_2O_3$ stack was fabricated as follows (See Extended Data Fig. 1 for the detailed fabrication procedure). First, a 330-nm-thick SiN layer was deposited on a thermally oxidized Si wafer by plasma enhanced chemical vapor deposition (PECVD). The thickness of the thermally oxidized $SiO_2$ layer was 4 μm. SiN waveguides were fabricated by electron-beam (EB) lithography and inductively coupled plasma (ICP) dry etching. Three HZO/$Al_2O_3$ multilayer stacks were then deposited on the Si waveguide by ALD. After depositing TiN on the HZO/$Al_2O_3$ stacks, annealing at 400 °C for 1 min was performed for achieving the ferroelectricity of HZO. TiN was removed by wet etching, and $SiO_2$ passivation was carried out by PECVD. The thickness of the $SiO_2$ cladding was 600 nm. Finally, Al electrodes were formed by metal sputtering and lift-off. For optical intensity and phase modulation measurement, a continuous-wave light from a tunable laser source (Santec, TSL-510) was injected into the SiN waveguide from a single-mode optical fiber through a SiN grating coupler. The polarization of the input light was tuned to the TE mode of the SiN waveguide using an in-line polarization controller. The output from the SiN waveguide was coupled back to the optical fiber through a grating coupler and measured by an InGaAs power meter (Agilent, 81624A). A precision source/measure unit (Agilent, B2902A) was used for the voltage application and the current measurement.

## Data availability

The data that support the findings of this study are available from the corresponding authors on reasonable request.

## Acknowledgements

This work was partly based on results obtained from projects (JPNP16007 received by M.T.) commissioned by the New Energy and Industrial Technology Development Organization (NEDO) and partly supported by JST, CREST (JPMJCR2004 received by M.T.), JST, MIRAI (JPMJMI20A1 received by M.T.), and “Advanced Research Infrastructure for Materials and Nanotechnology in Japan (ARIM)” of the Ministry of Education, Culture, Sports, Science and Technology (MEXT) (JPMXP1222UT1028).

## Author contributions

K.Ta. contributed to fabrication, measurement, and manuscript preparation, N.S. contributed to the idea, K.W. contributed to fabrication and measurement, Y.M. contributed to design and measurement, T.A. contributed to manuscript preparation, and H.S. contributed to simulation. K.To and S.T contributed to the overall discussion. M.T. contributed to the idea, discussion, and manuscript revision and also provided high-level project supervision.

**Competing interests**

The authors declare no competing interests.

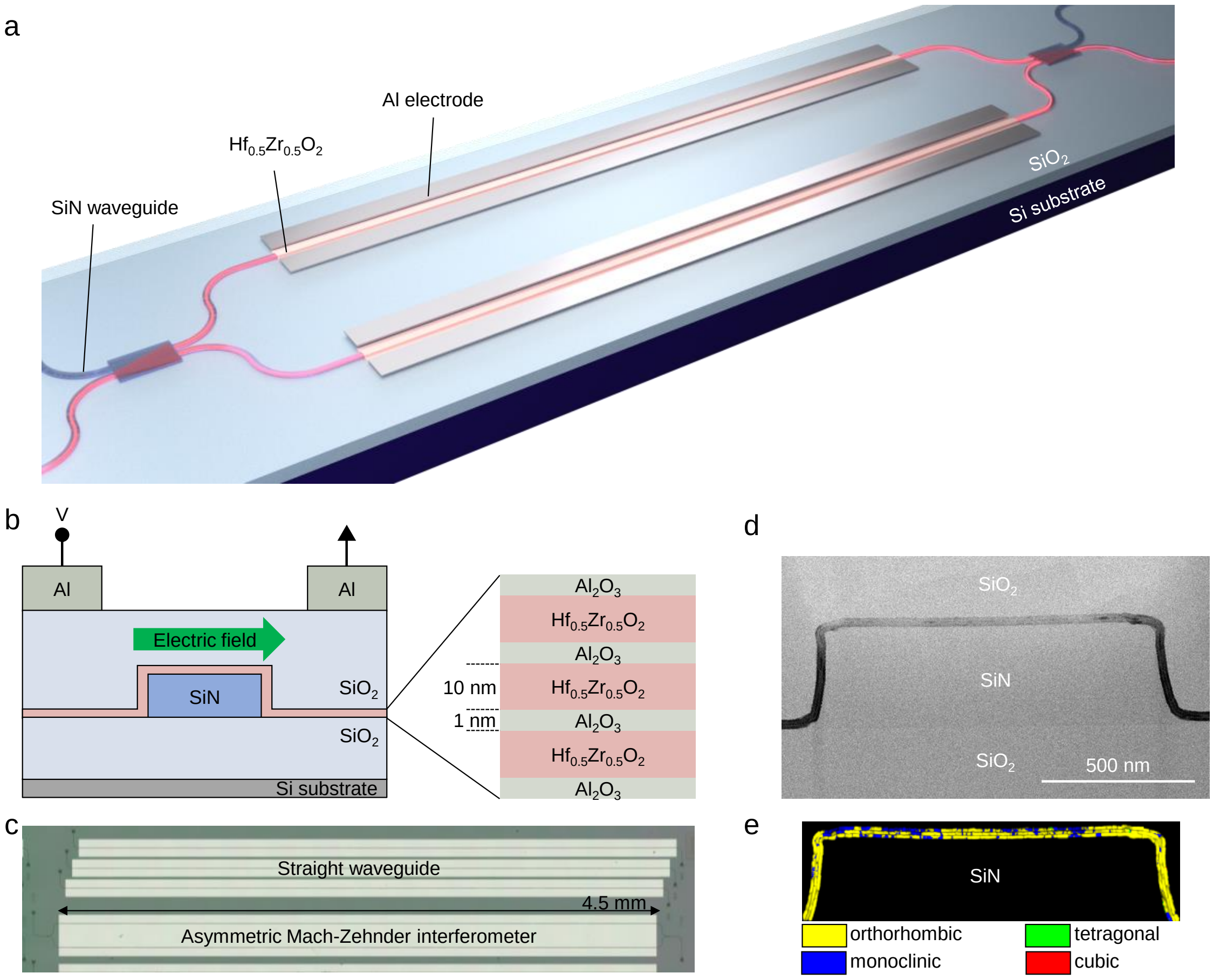

Fig. 1. **Schematic and images of a SiN waveguide with $Hf_{0.5}Zr_{0.5}O_2$-based optical phase shifter. a,** 3D view of a Mach-Zehnder interferometer based on a SiN waveguide with $Hf_{0.5}Zr_{0.5}O_2$-based optical phase shifters. Al electrodes are formed to apply an electric field to $Hf_{0.5}Zr_{0.5}O_2$. **b,** Cross-sectional structure of the optical phase shifter with $Hf_{0.5}Zr_{0.5}O_2/Al_2O_3$ stacks deposited on a SiN waveguide. An external electric field is applied along the transverse direction through the Al electrodes. **c,** Plan-view microscopy image of the fabricated device. An asymmetric Mach-Zehnder interferometers and straight waveguides with grating couplers are prepared for optical phase and loss evaluation. **d,** Cross-sectional TEM image of the SiN waveguide with $Hf_{0.5}Zr_{0.5}O_2/Al_2O_3$ stacks. **e,** Phase map of the $Hf_{0.5}Zr_{0.5}O_2$ layers.

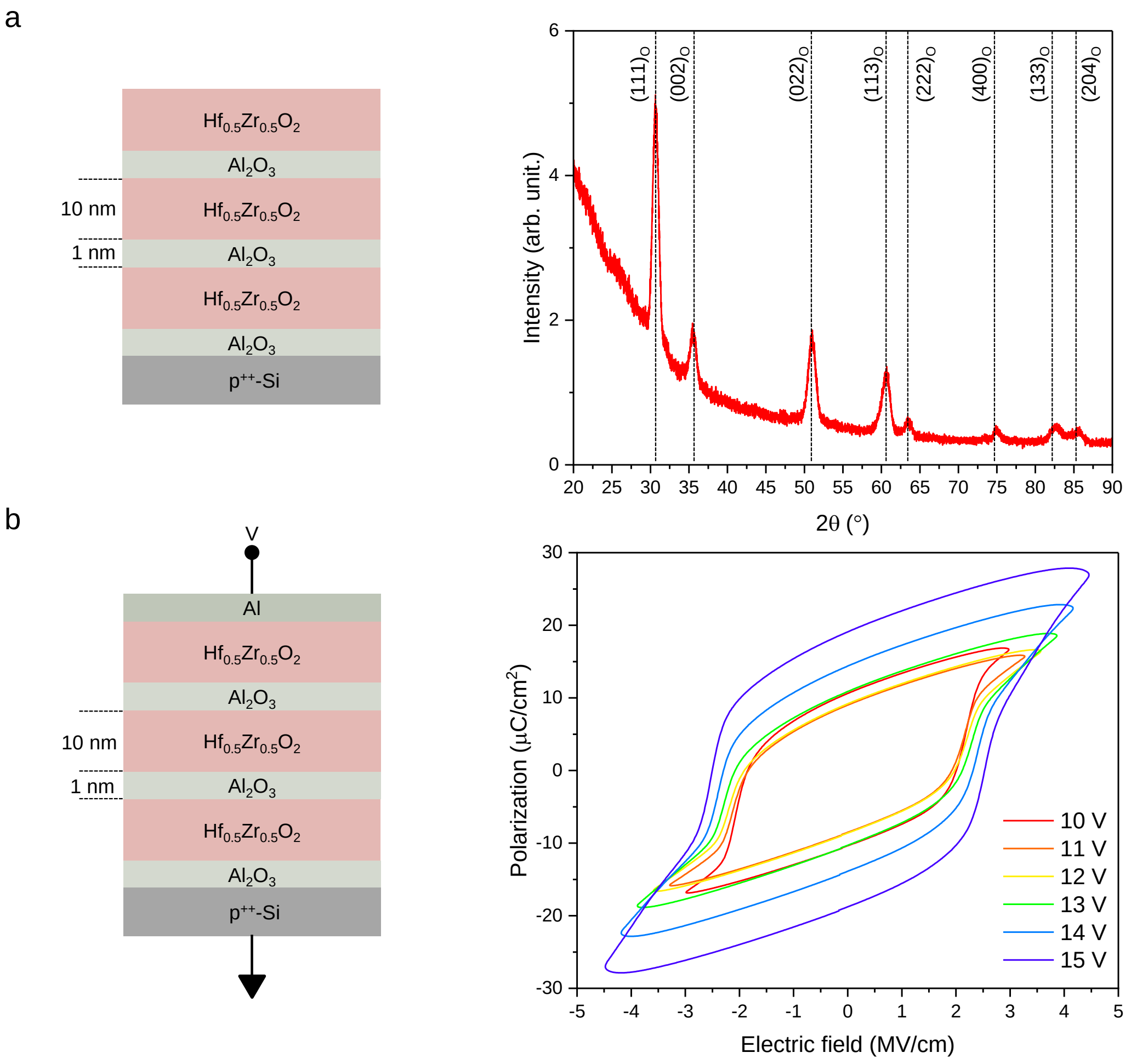

Fig. 2. **Material characterizations of $Hf_{0.5}Zr_{0.5}O_2/Al_2O_3$ stacks. a,** Grazing-incidence X-ray diffraction of the $Hf_{0.5}Zr_{0.5}O_2/Al_2O_3$ stacks deposited on a Si substrate. The orthorhombic phase of the $Hf_{0.5}Zr_{0.5}O_2$ layer is confirmed by the diffraction peaks. The diffraction peaks **b,** Polarization−voltage (PV) characteristics of the metal-insulator-semiconductor capacitor composed of the $Hf_{0.5}Zr_{0.5}O_2/Al_2O_3$ stacks. The hysteresis PV curves clearly indicates the ferroelectricity of the $Hf_{0.5}Zr_{0.5}O_2$ layer.

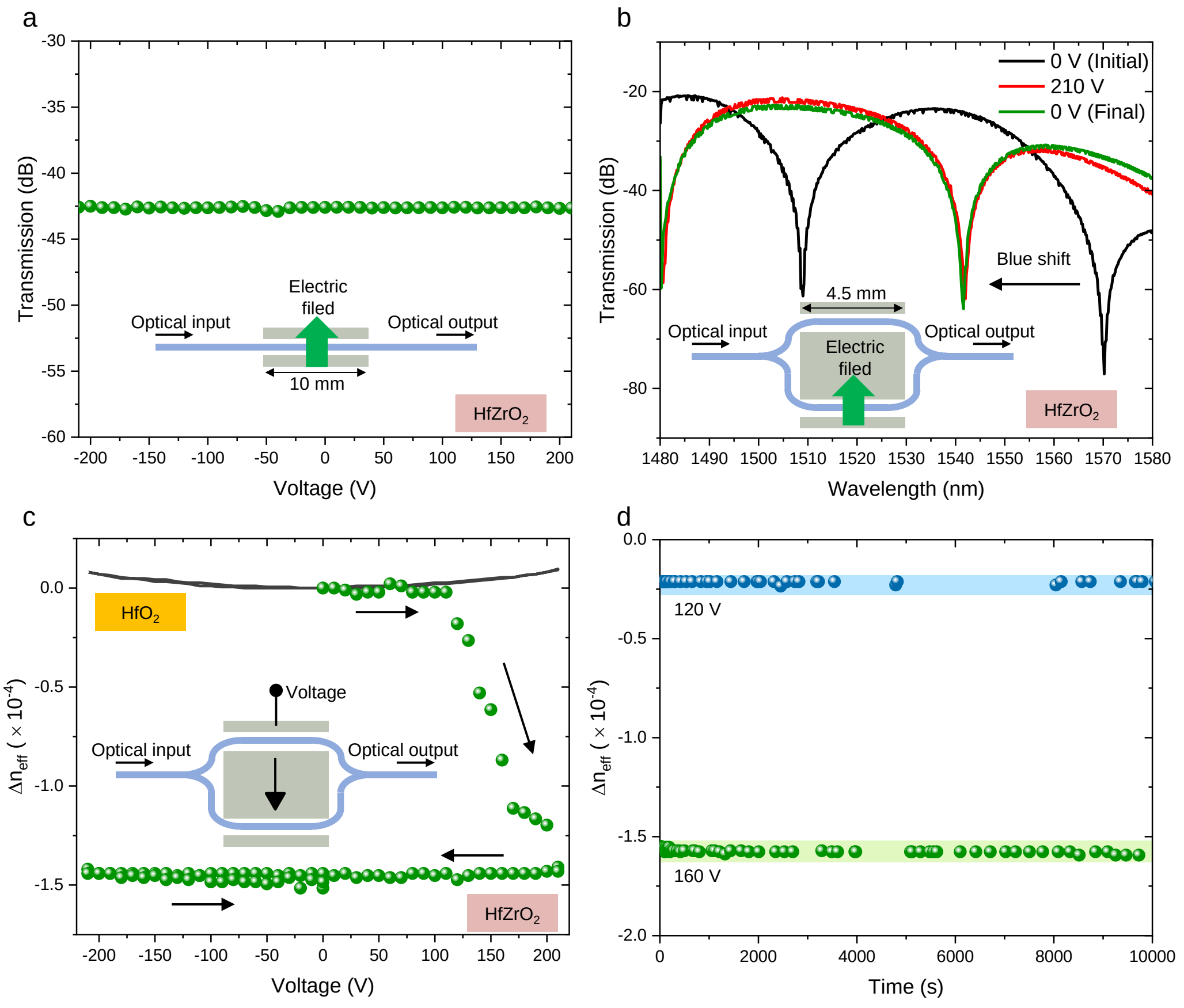

Fig. 3. **Optical properties of $Hf_{0.5}Zr_{0.5}O_2$-based optical phase shifter. a,** Optical transmission measurement of the 10-mm-long straight device with respect to the voltage application. No significant intensity modulation was observed, indicating that the free-carrier effects are negligible. **b,** Transmission spectra of the Mach-Zehnder interferometer with the 4.5-mm-long phase shifter. A blue shift in the resonance wavelength peak is observed when 210 V is applied once, showing the non-volatile optical phase shift. **c,** The change in the effective refractive index when a voltage is swept between +210 and -210 V. When a voltage increases from 0 V to 210, the significant negative refractive index change is observed. Following the initial significant refractive index change, only a small positive change is observed, which can be attributed to the Kerr effect in SiN. A similar Kerr effect is also observed in the $HfO_2$ and $SiO_2$ devices. **d,** The retention characteristics

of the non-volatile refractive index change after a voltage application of 120 V or 160 V. No significant change in the refractive index change is observed up to $10^4$ s.

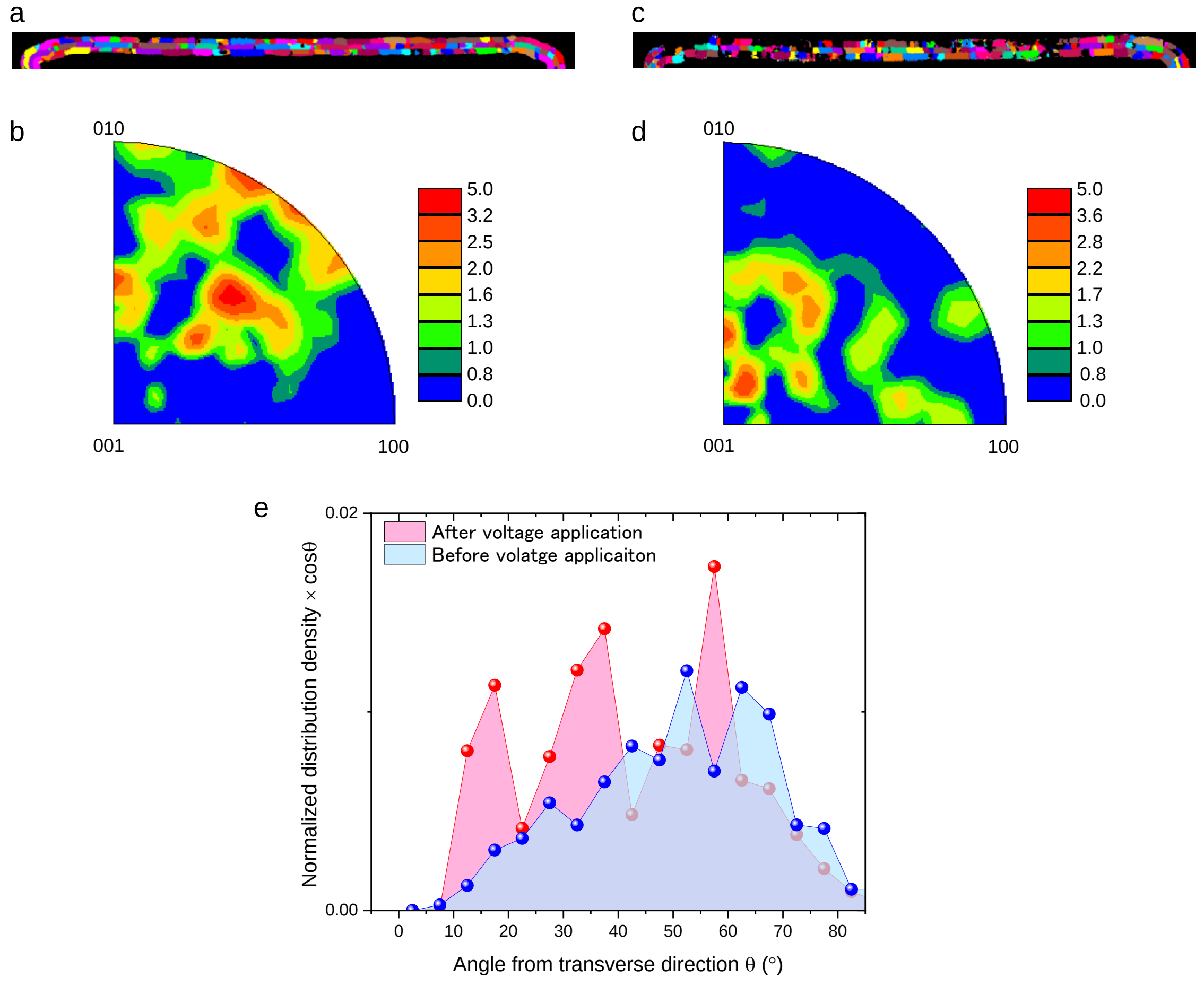

Fig. 4. **Automated crystal orientation mapping of the $Hf_{0.5}Zr_{0.5}O_2$ layers on the top of the SiN waveguide. a, b,** Grain map and inverse pole figure of the orthorhombic $Hf_{0.5}Zr_{0.5}O_2$ layer before voltage application. **c, d,** Grain map and inverse pole figure of the orthorhombic $Hf_{0.5}Zr_{0.5}O_2$ layer after voltage application. **e,** The product of the normalized distribution density of the orientation angle θ of the orthorhombic phase and cosθ. The polarization axis rotates toward the direction of an external electric field due to the voltage application, which contributes to the change in refractive index.

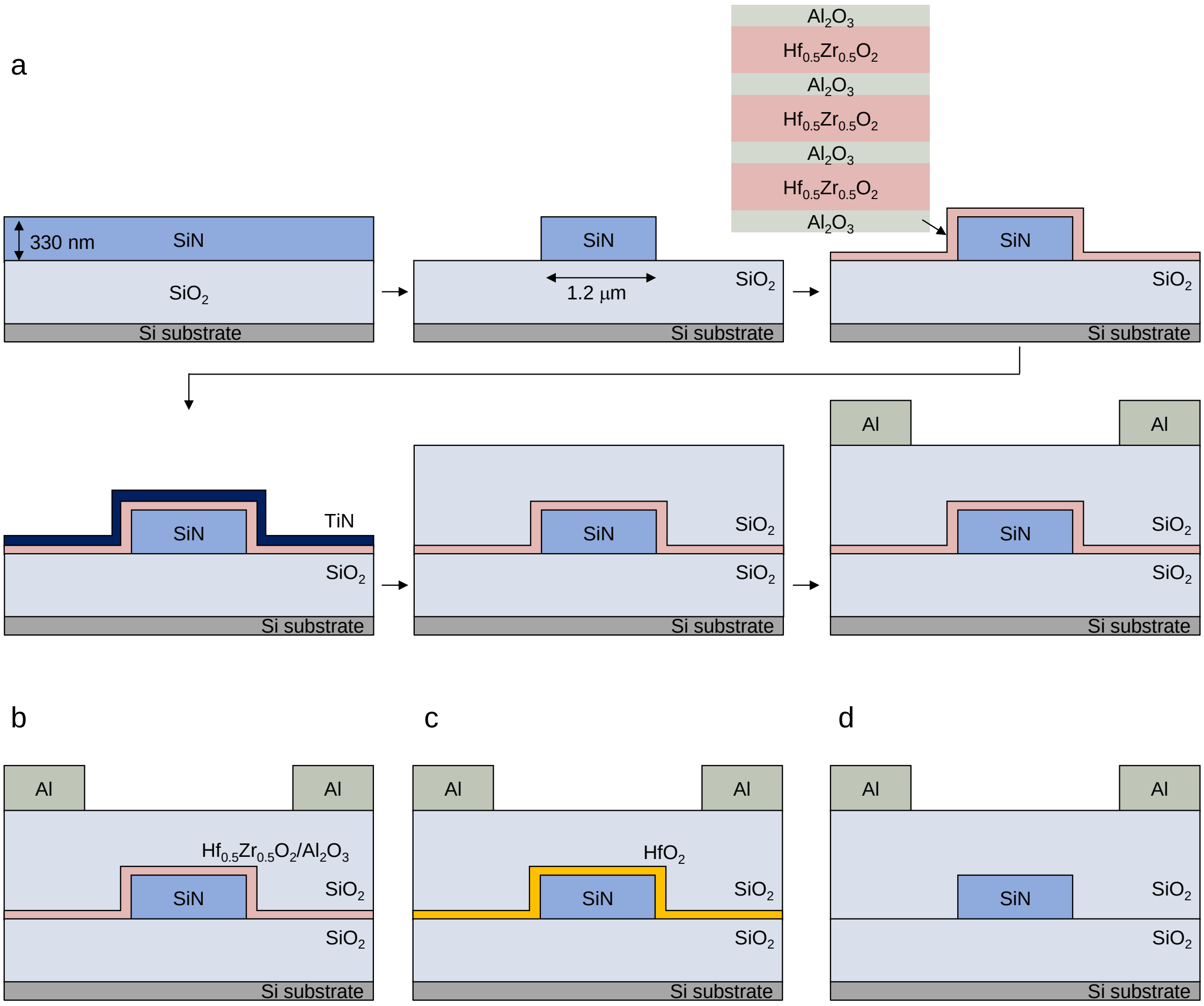

Extended Data Fig. 1. **Device fabrication. a,** Fabrication procedure of the optical phase shifter with $Hf_{0.5}Zr_{0.5}O_2/Al_2O_3$ stacks deposited on the SiN waveguide. After depositing a SiN layer on a thermally oxidized Si wafer, SiN waveguides are formed. After depositing $Hf_{0.5}Zr_{0.5}O_2/Al_2O_3$ stacks by atomic layer deposition, TiN is deposited followed by annealing at 400 °C for 1 min. After removing TiN, $SiO_2$ cladding and Al electrodes are deposited. **b, c, d,** Cross-sectional schematics of the SiN waveguide with the $Hf_{0.5}Zr_{0.5}O_2/Al_2O_3$ stacks, $HfO_2$, and $SiO_2$, respectively.

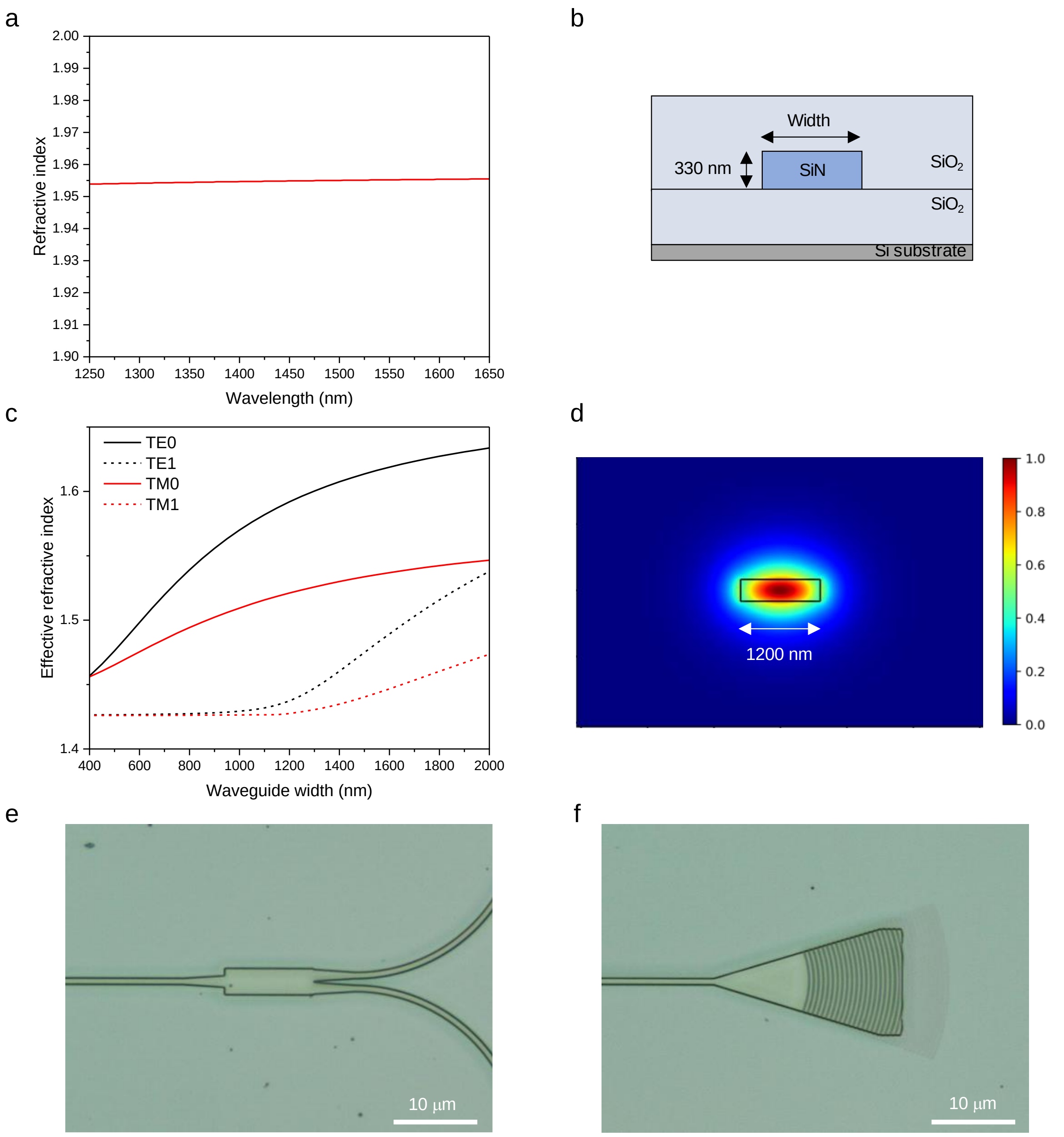

Extended Data Fig. 2. **Analysis of SiN waveguide and plan-view images of SiN waveguide components. a,** Refractive index of the SiN layer measured by spectroscopic ellipsometry. **b,** Cross-sectional schematic of a SiN waveguide for numerical analysis. **c,** Effective refractive index of the fundamental and first TE and TM modes as a function of the waveguide width. **d,** Mode profile of the fundamental TE mode when the waveguide width is 1200 nm. **e,** Plan-view microscopy image of the MMI coupler for AMZI. **f,** Plan-view microscopy image of the grating coupler.

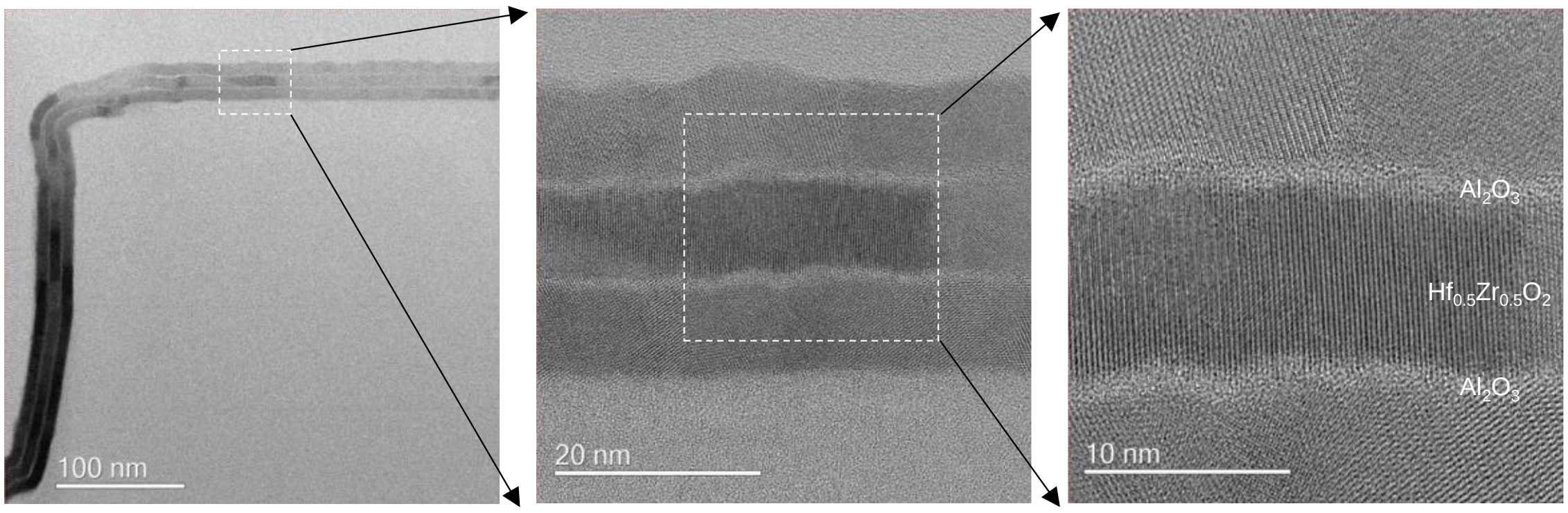

Extended Data Fig. 3. **Cross-sectional TEM image of $Hf_{0.5}Zr_{0.5}O_2/Al_2O_3$ stacks deposited on the SiN waveguide.** Three polycrystalline $HfZrO_2$ layers separated by $Al_2O_3$ interlayers are clearly observed.

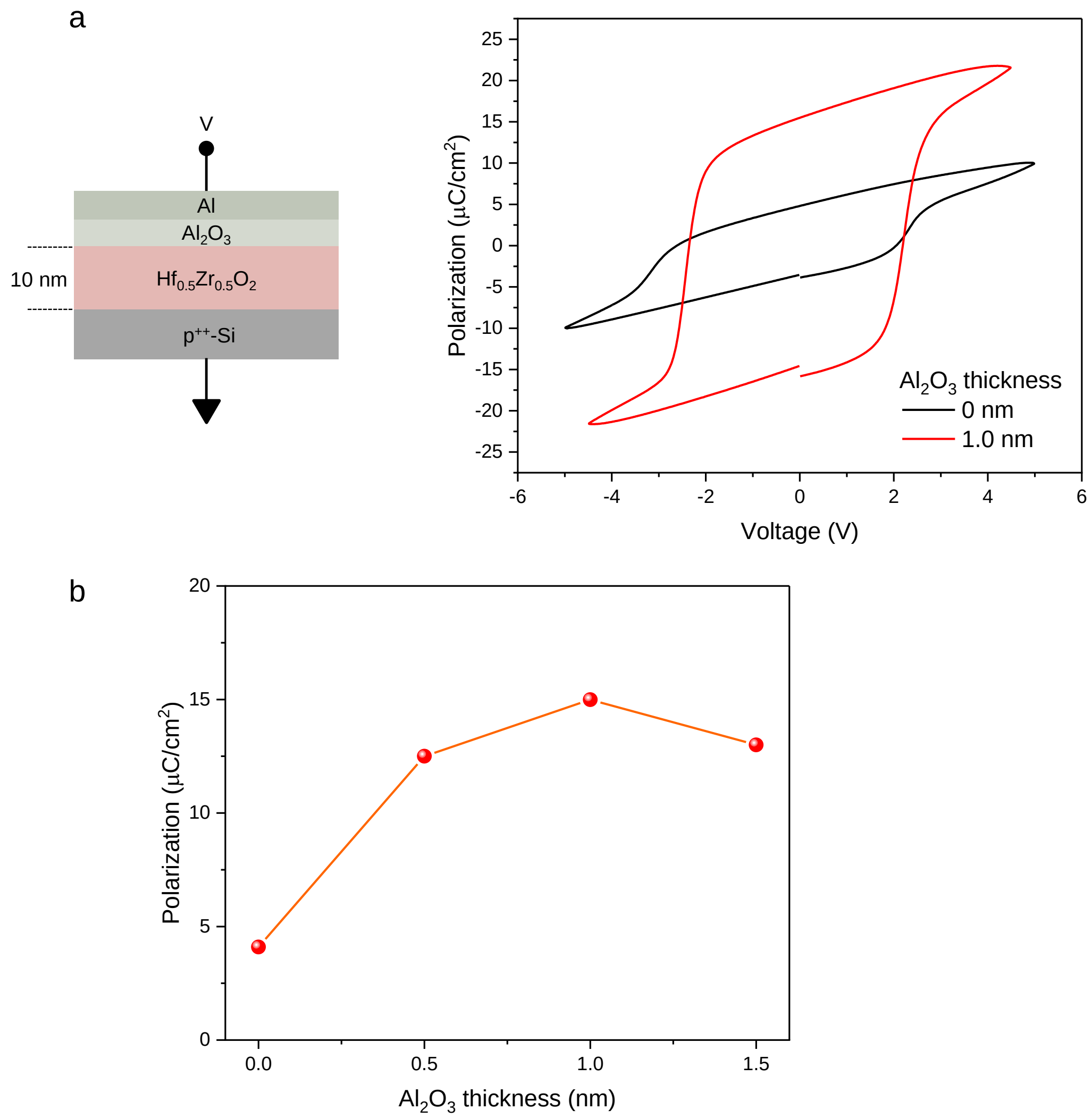

Extended Data Fig. 4. **Ferroelectric properties of $Hf_{0.5}Zr_{0.5}O_2$ with $Al_2O_3$ capping layer. a,** Polarization–voltage (PV) characteristics of metal-insulator-semiconductor capacitors composed of a 10-nm-thick single-layer $Hf_{0.5}Zr_{0.5}O_2$ fabricated on a heavily-doped p-type Si substrate. **b,** The remanent polarization as a function of $Al_2O_3$ thickness. As the thickness of the $Al_2O_3$ capping layer increases, the remanent polarization increases and saturates with a 1-nm-thick $Al_2O_3$.

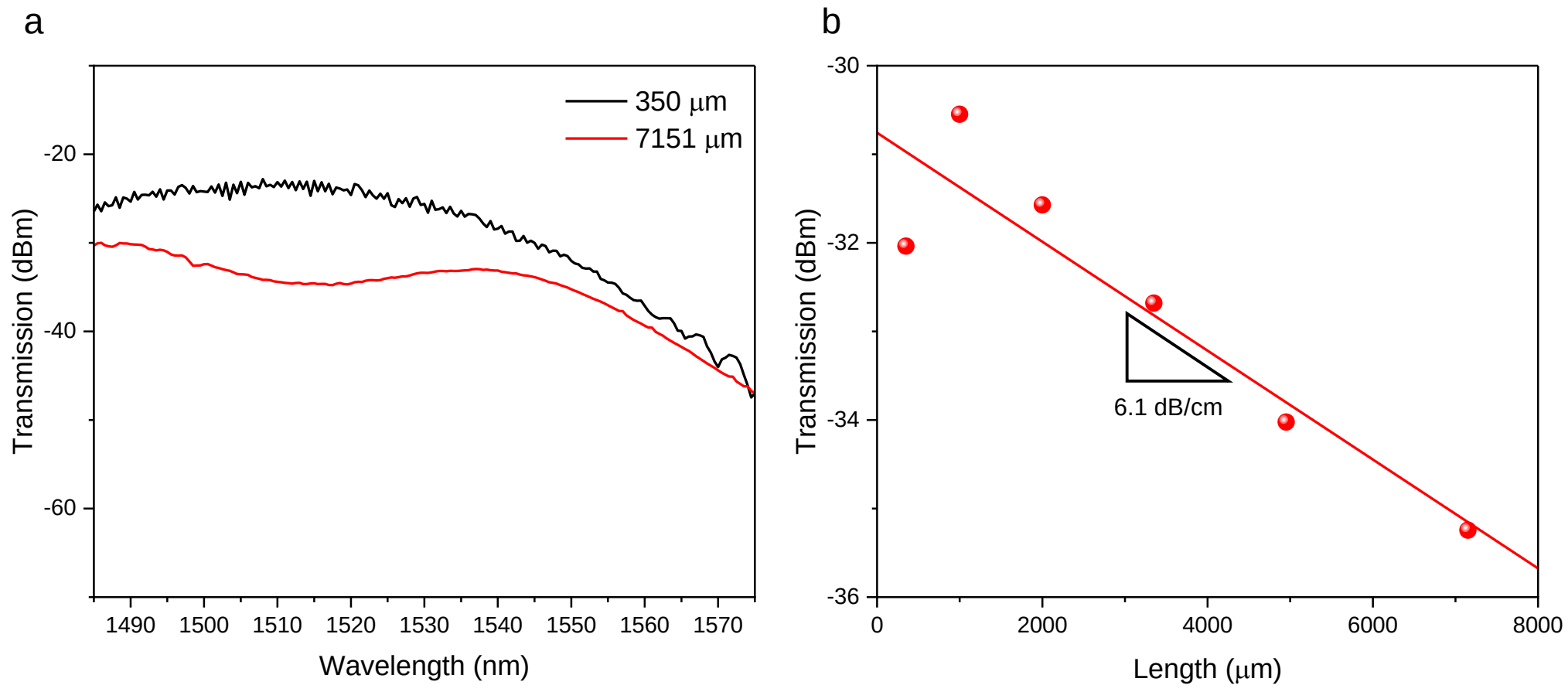

Extended Data Fig. 5. **Transmission properties of straight SiN waveguides with grating couplers. a,** Transmission spectra of the SiN waveguides with lengths of 350 μm and 7151 μm. The observed attenuation in the wavelength range of 1500 – 1530 nm for the 7151-μm-long device is attributed to the absorption caused by the N-H bonds in SiN. **b,** The transmission at a wavelength of 1550 nm as a function of the device length. The propagation loss is found to be 6.1 dB/cm.

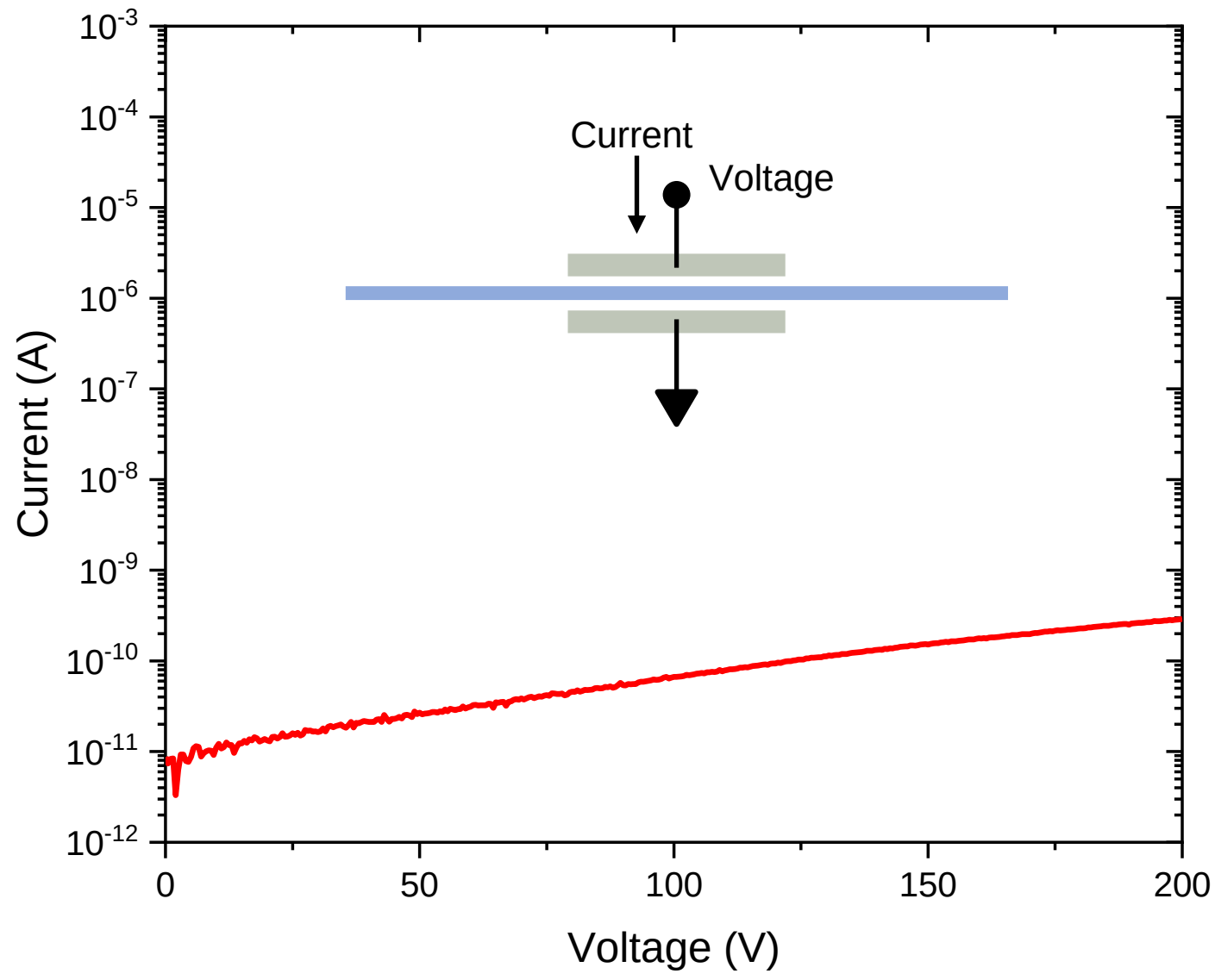

Extended Data Fig. 6. **Current-voltage curve of the 10-nm-long optical phase shifter.** The leakage current is less than 300 pA even at 200 V, suggesting that there is no significant thermo-optic effect during the measurement of the phase shift.

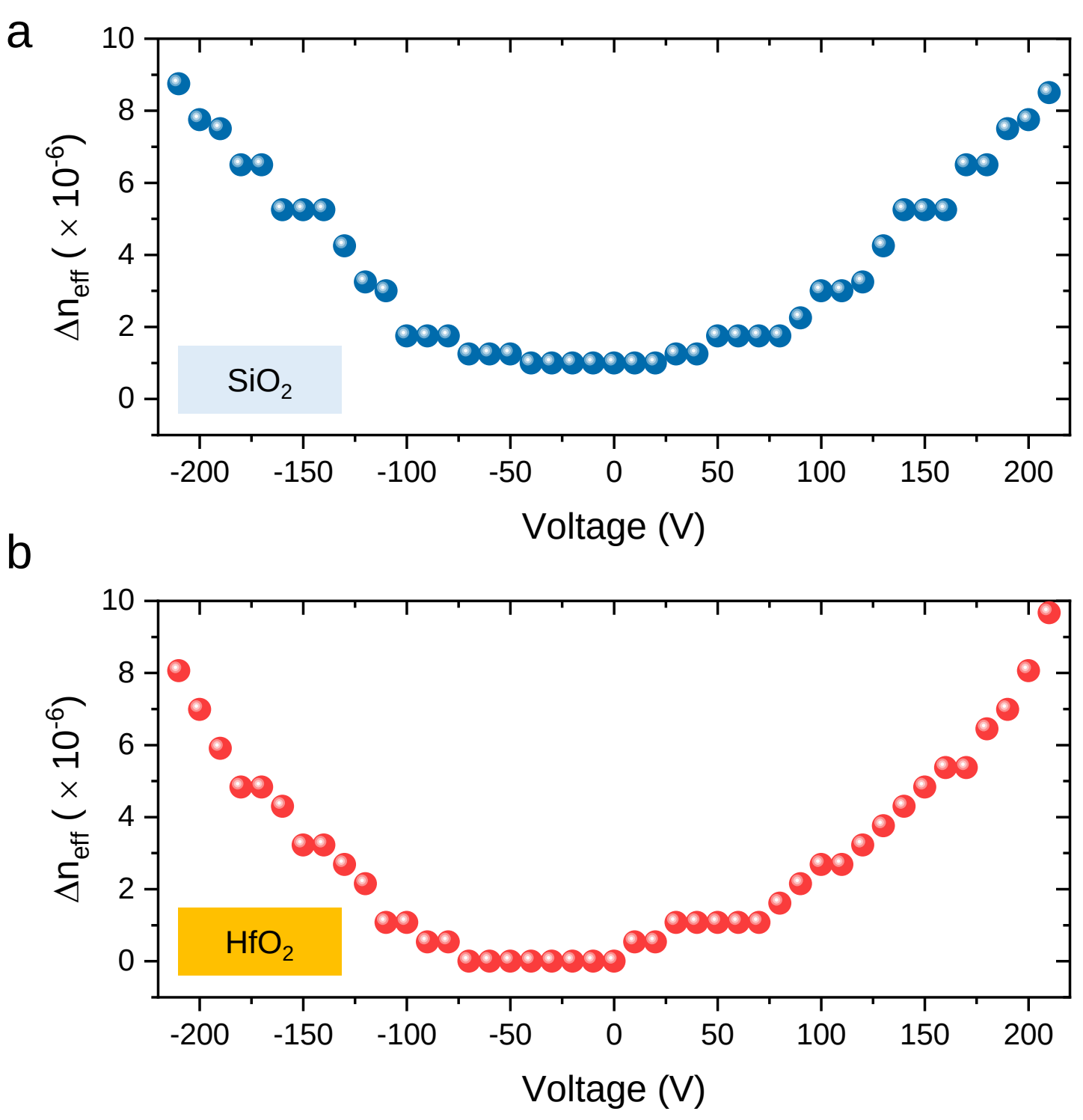

Extended Data Fig. 7. **The change in the effective refractive index in the $SiO_2$ and $HfO_2$ devices when a voltage is swept between +210 and -210 V. a,** The change in the effective refractive index of the $SiO_2$ device. **b,** The change in the effective refractive index of the $HfO_2$ device. Both $SiO_2$ and $HfO_2$ exhibits the positive refractive index change that is proportional to the square of the applied voltage. The refractive index change can be attributed to the Kerr effect in SiN.

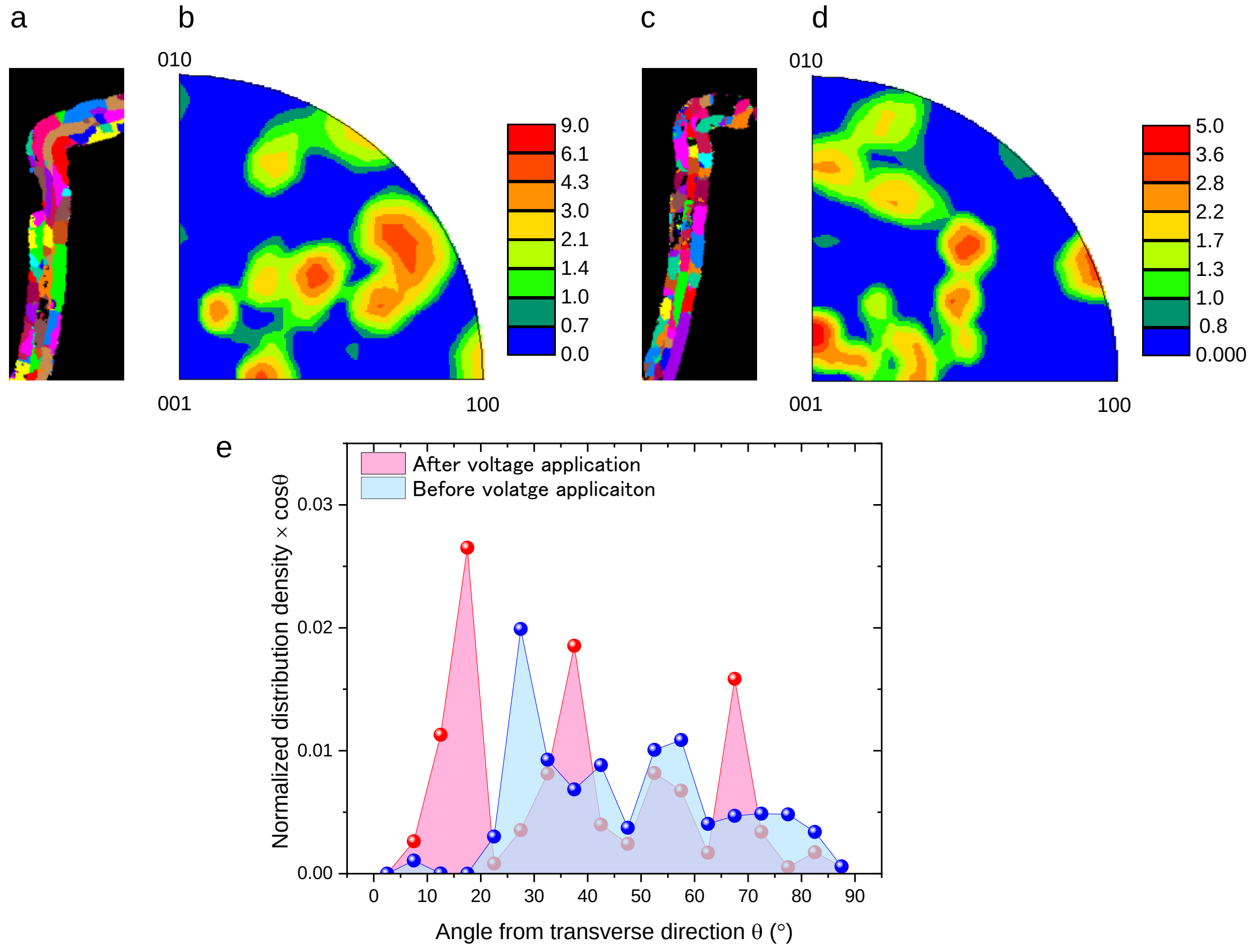

Extended Data Fig. 8. **Automated crystal orientation mapping of the $Hf_{0.5}Zr_{0.5}O_2$ layers at the sidewall of the SiN waveguide. a, b,** Grain map and inverse pole figure of the orthorhombic $Hf_{0.5}Zr_{0.5}O_2$ layer before voltage application. **c, d,** Grain map and inverse pole figure of the orthorhombic $Hf_{0.5}Zr_{0.5}O_2$ layer after voltage application. **e,** The product of the normalized distribution density of the orientation angle θ of the orthorhombic phase and cosθ. The polarization axis rotates toward the direction of an external electric field due to the voltage application, which contributes to the change in refractive index. The total area under the distribution, corresponding to the average of the transverse component, increased by approximately 20% after the voltage application.